\documentclass[reprint, amsmath,
superscriptaddress,
nobibnotes,
nofootinbib,showpacs,
amssymb,aps,prb]{revtex4}

\usepackage[dvips]{graphicx}% Include figure files
\usepackage{dcolumn}% Align table columns on decimal point
\usepackage{bm}% bold math

\newcommand{\gh}{{\it graphane}}
\newcommand{\Gh}{{\it Graphane}}
\newcommand{\ssh}{SSHGraphene}
\newcommand{\etal}{{\it et al}}

\long\def\symbolfootnote[#1]#2{\begingroup%
\def\thefootnote{\fnsymbol{footnote}}\footnote[#1]{#2}\endgroup}

\begin{document}

\title{Single-side-hydrogenated graphene: Density functional theory predictions}
\author{Bhalchandra S. Pujari}
\affiliation{National Institute for Nanotechnology, 11421 Saskatchewan Drive, Edmonton, AB, T6G 2M9, Canada}
\affiliation{Department of Mechanical Engineering, University of Alberta, Edmonton, AB, T6G 2G8, Canada}
\author{Sergey Gusarov}
\affiliation{National Institute for Nanotechnology, 11421 Saskatchewan Drive, Edmonton, AB, T6G 2M9, Canada}
\author{Michael Brett}
\affiliation{National Institute for Nanotechnology, 11421 Saskatchewan Drive, Edmonton, AB, T6G 2M9, Canada}
\affiliation{Department of Electrical and Computer Engineering, University of Alberta, Edmonton AB, T6G 2V4, Canada}
\author{Andriy Kovalenko}
\email[Corresponding Author: ]{andriy.kovalenko@nrc-cnrc.gc.ca}
\affiliation{National Institute for Nanotechnology, 11421 Saskatchewan Drive, Edmonton, AB, T6G 2M9, Canada}
\affiliation{Department of Mechanical Engineering, University of Alberta, Edmonton, AB, T6G 2G8, Canada}

\begin{abstract}
Hydrogenation has proven to be an effective tool to open the bandgap of graphene.
In the present density functional study we demonstrate that single-side-hydrogenated
graphene is a semiconductor with an indirect bandgap of 1.89 eV, in between the gapless
graphene and wide bandgap {\it graphane}. We show that its electronic structure and lattice
characteristics are substantially different from those of graphene, graphone, or {\it graphane}.
The lattice parameter and C-C bond length are found to be lengthened by 15\% of those
of graphene. Our binding energy analysis confirms that such a single sided hydrogenation
leads to thermodynamically stable material.
\end{abstract}

% PACS, the Physics and Astronomy

\maketitle

Graphene is unarguably one of the most novel materials of modern time. Graphene has
potential application in a wide range of fields,\cite{geim:nat} but the lack of bandgap is
one obstacle for utilizing it in semiconductor based devices. There are a number of
investigations on bandgap opening and tuning using various techniques. Giovannetti \etal
\cite{PhysRevB.76.073103} had predicted theoretically that the substrate might induce a
small bandgap. However, experimentally substrate induced bandgap is an issue of
controversy.\cite{Zhou2007770,Rotenberg2008258,Zhou2} A more accepted approach is by
creating finite size nanoribbons out of graphene. The electronic confinement in
nanoribbons results in a substantial bandgap opening.\cite{Berger26052006,Han2007,Barone}
Controlling the width of a nanoribbon is a technologically challenging task, which puts a
restriction on this method for precise tuning. A more promising method is hydrogenation
of graphene.

The interest in hydrogenation of graphene is twofold; it can be used
to tune the bandgap and it can also be harnessed as a hydrogen
storage material. In an important theoretical work, Duplock \etal
\cite{Duplock2004} first showed that upon adsorption of hydrogen on graphene a
bandgap can be opened. Even a single hydrogen atom may lead to bandgap
of 1.25 eV. Later, Boukhvalov {\etal}\cite{PhysRevB.77.035427} extensively
studied the adsorption of hydrogen atoms on multiple layers of graphene. Their
density functional study brought out the nature of the magnetism as well as
stability of graphene in the presence of adsorbents. More recently, in an
intriguing experiment Balog \etal \cite{balog} demonstrated that the
hydrogenation can indeed be used to open the bandgap. Their results confirmed
that patterned hydrogenation can be systematically applied to open up a
bandgap. By a novel way of semi-hydrogenating graphene, Zhou {\etal}
\cite{zhou_graphone} proposed a new ferromagnetic material, graphone, which
exhibits a small indirect bandgap (0.46 eV). As we shall demonstrate in this
paper, complete hydrogenation eventually leads to an indirect but a larger band gap.

All of the above reports deal with partial hydrogenation of graphene. In a different
approach, Sofo {\etal} \cite{sofo:153401} in 2007 predicted the existence of completely
hydrogenated graphene and named it {\gh}.\symbolfootnote[2]{`{\Gh}' is
written in italic to avoid confusion with `graphene'.} It was two years later that
Elias {\etal}\cite{elias01302009} experimentally confirmed the existence of the same,
which sparked tremendous interest.\cite{pujari:graphane,
Srinivasan2009, barnard, PhysRevB.81.205417, Flores2009} Recently, Chandrachud {\etal}
\cite{prachi} carried out a systematic investigation of hydrogenation of graphene. They
have analyzed the transition from graphene to {\gh} as a function of hydrogenation and
confirmed that the amount of hydrogenation determines the bandgap.

Although hydrogenation of graphene has been under a series of investigations, the
research addressed either partial hydrogenation (as in graphone) or complete but double
sided hydrogenation (as in \gh). A recent report by Muniz and Maroudas \cite{muniz:113532}
considers the effects of hydrogen coverage on a single side of graphene plane. However,
their study is based on the classical methods and does not address the electronic structure.
At present, there is a compelling necessity to study in detail the electronic structure of
single sided hydrogenation of graphene. It stems from the fact that when graphene is grown on
a substrate like SiO$_2$ it is impossible for hydrogen to diffuse through the graphene to form
double sided hydrogenation\cite{elias01302009,bunch} and the only possibility is partial or
complete hydrogenation from a single side. Neglecting the substrate effects on graphene which 
were found to be relatively small,\cite{PhysRevB.76.073103,Zhou2007770,Rotenberg2008258,Zhou2}
we concentrate in this report on the electronic structure properties characterizing such
single-side-hydrogenated graphene, which we shall call {\ssh}.

Our results bring out some remarkable properties of this material such as the indirect
bandgap in its band structure. We also analyze the density of states (DOS) and their
angular momentum decomposed counterparts along with the orbital charge density,
in order to exhibit the nature of its electronic structure.

We carried out full unit cell optimization and detailed electronic structure calculations 
of {\ssh} using density functional theory (DFT) with a plane wave basis as implemented in 
the Quantum ESPRESSO package\cite{qe}. We used ultrasoft pseudopotentials \cite{Vanderbilt} 
and the Generalized Gradient Approximation (GGA) with the exchange-correlation functional 
of Perdew, Burke, and Ernzerhof (PBE).\cite{PBE} The plane wave basis cut-off was 60 Ry 
($\sim$ 816 eV) for wave functions and 240 Ry ($\sim$ 3265 eV) for charge density, 
sufficiently large for graphene-like systems. A convergence threshold was set at 
10$^{-8}$ eV for the energy self-consistency and 0.005 eV/{\AA} for the forces. 
Further, to maintain the accuracy, integration over the Brillouin zone was performed 
on regular $26 \times 26 \times 1$ Monkhorst-Pack grids. The band structure was plotted 
on the lines joining the $M$, $\Gamma$, $K$, and $M$ points, and the individual line 
segments were sampled using 50 grid points each. The corresponding precision was also 
maintained for the cell optimization carried out using the BFGS quasi-Newton algorithm. 
The convergence threshold on the pressure was kept at 0.1 kBar. The computational unit 
cell consisted of two carbons and two hydrogens. A vacuum space of 12 {\AA} was kept 
normal to the (\ssh) plane to avoid any interactions between the adjacent sheets. 
We also validated our results for the bandgaps, bond lengths, and lattice parameters 
by making a comparison of different exchange-correlation functionals (see the Supplementary 
Material).

\begin{figure}[ht]
\begin{center} 
\includegraphics[width=7cm]{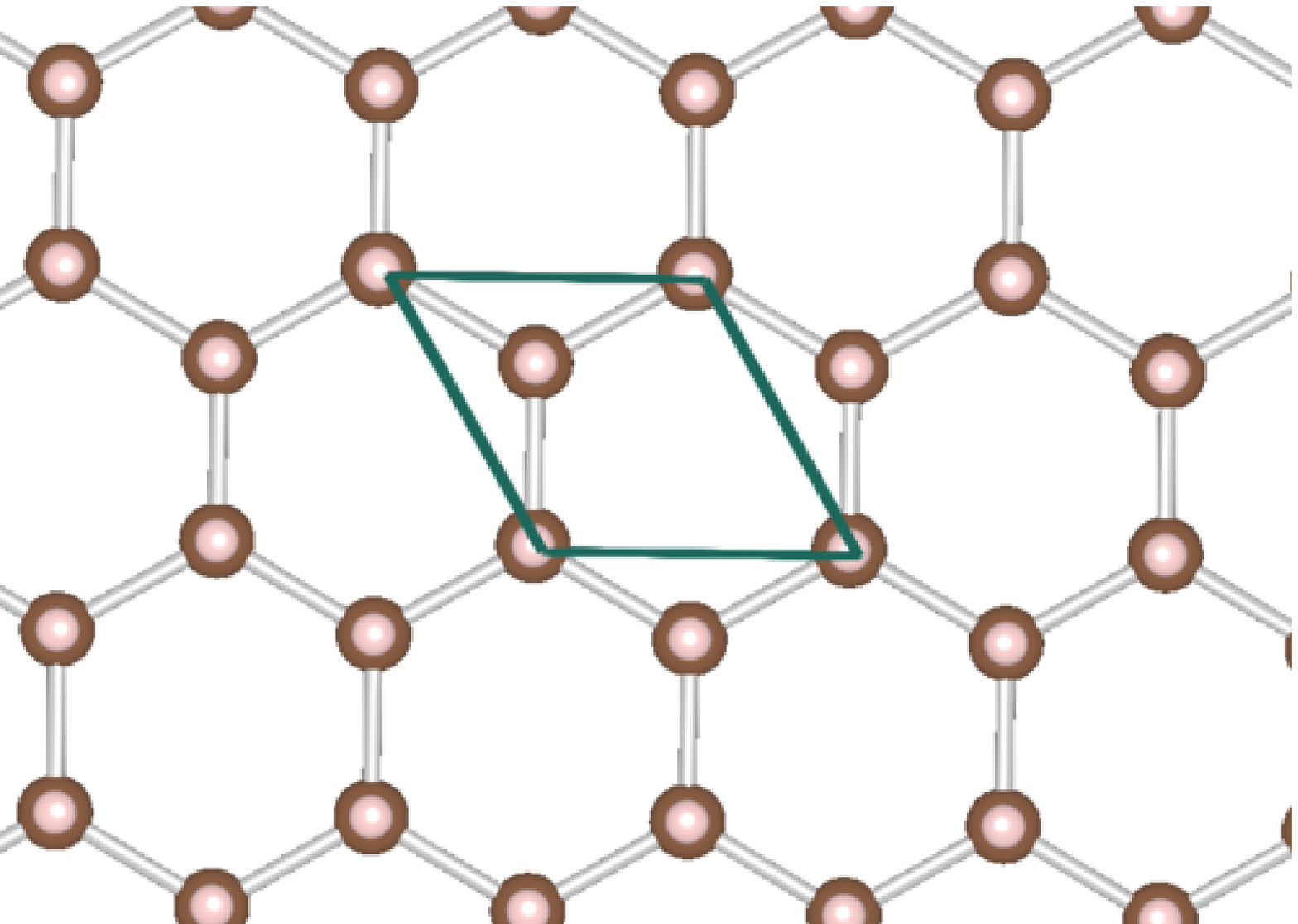}
\hskip 0.5cm
\includegraphics[width=7cm]{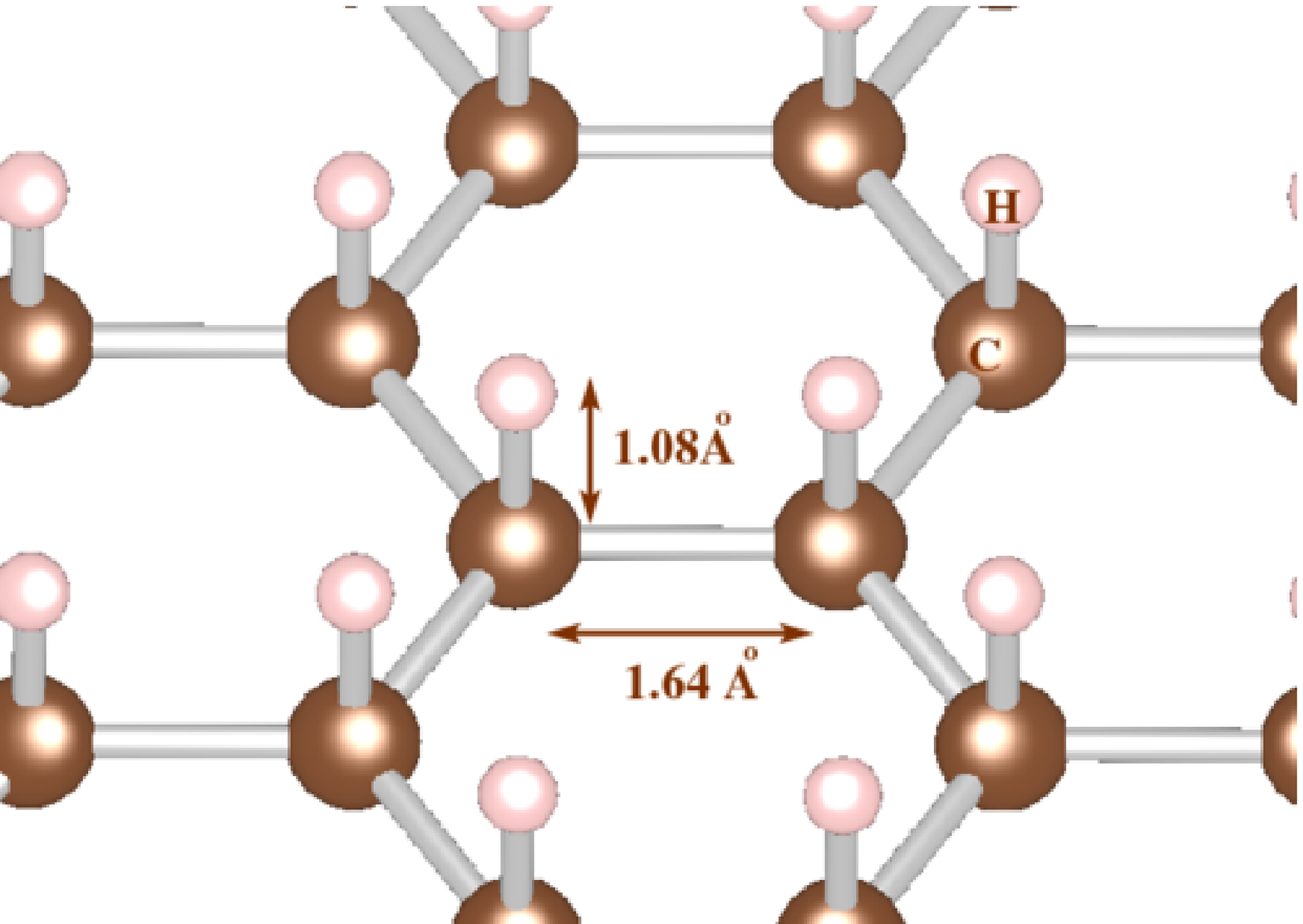}
\end{center}
\caption{
Top view of the {\ssh} hexagonal structure (the solid line rhombus marks the unit cell) 
with carbon and hydrogen atoms shown in dark and lighter shade, respectively. 
A close-up with the bond lengths in Angstr\"{o}ms.}
\label{fig:geom}
\end{figure}

We first discuss the geometry and then the electronic structure of
single-side-hydrogenated graphene ({\ssh}). Whenever possible we shall put our results
in perspective of graphene and \gh. For such a comparison it is worthwhile to review
some properties of graphene and {\gh} before we discuss {\ssh}.

Graphene is a one-atom-thick sheet of sp$^2$-bonded carbon atoms that are densely packed
in a bipartite crystal lattice. It has two atoms per unit cell which has the lattice
parameter of 2.46 \AA, with a carbon-carbon bond length of 1.42 \AA. Although {\gh} is
bipartite and hexagonal, its unit cell has four atoms (two carbons and two hydrogens) 
and has a larger lattice parameter, namely 2.51 \AA.\cite{sofo:153401} In {\gh} every
alternate carbon atom is attached to a hydrogen atom from alternate sides of the plane.
In response to the addition of hydrogens, the carbon atoms are displaced out of the plane
towards hydrogen atoms. In short, the carbon atoms in {\gh} are no longer planar.

The unit cell of {\ssh} also contains four atoms, two carbons and two hydrogens. 
We carried out full optimization of the unit cell, including both the unit cell geometry and 
the atomic positions. The optimized geometry of {\ssh} is shown in Figure \ref{fig:geom} 
depicting a top view with the unit cell marked with the solid line and a close-up with 
the bond lengths. As seen from the figure the cell is similar to that of graphene, except 
that the lattice parameter for {\ssh} is now enlarged to 2.83 {\AA}, which is larger than 
{\gh} (2.51 \AA) as well. Notice that the enhancement is necessary in order to accommodate 
the hydrogen atoms,as the unoptimized unit cell of graphene does not favor the complete 
hydrogenation. The increase in the lattice parameter is due to the increase in the 
carbon-carbon bonds.As seen from the zoomed-in view, the carbon-carbon bond length is 
increased from 1.42 {\AA} (in graphene) to 1.64 \AA. The lattice is thus expanded by 15\%. 
The increase in the bond length upon hydrogenation is not surprising, as the same effect 
is also seen in {\gh}. Similarly, as expected, upon single sided hydrogenation the carbon
atoms remain in one plane with the hydrogens forming another plane at 1.08 \AA. This is a
typical bond length of C-H when bonded covalently. (In methane, for example, the bond
length is 1.09 \AA.) To summarize, a comparison of (available) structural parameters of
graphene, graphone, {\gh}, and {\ssh} are given in table \ref{tab:struct}. It also shows
the binding energy per atom which is the signature of thermodynamic stability of the system.
The biding energy for {\ssh} is calculated using the pseudoatomic energies of carbon ($E_{\rm C}$)
and hydrogen ($E_{\rm H}$) atoms, and using 
  $\Delta E = E_{\rm C} + E_{\rm H} - E_{\rm {\ssh}}$,
where $E_{\rm {\ssh}}$ is the total energy of {\ssh}. Thus, the higher the energy the 
more stable the system. The binding energies for graphene and {\gh} are as reported 
in the literature.\cite{sofo:153401} The overall trend is quite straightforward. 
Graphene, having the smallest C-C bond, is the most stable of all. Although not as stable 
as others,{\ssh} is still strongly bound. To put it in perspective, recall that benzene 
has the binding energy 6.49 eV/atom while acetylene has 5.90 eV/atom \cite{sofo:153401},
and both are among the most stable hydrocarbons. Thus there is no doubt that {\ssh}
is indeed thermodynamically very stable.

\begin{table}[tb]
\begin{tabular}{c|c|c|c|c}
	\hline
	~& Graphene & Graphone \cite{zhou_graphone}& {\Gh}\cite{sofo:153401} & {\ssh}\\
	\hline
	\hline
	$a$ (\AA) 	& 2.46 & -- & 2.51 	& 2.83\\
	C-C (\AA) 	& 1.42 & 1.50	& 1.52 	& 1.64\\
	C-H (\AA)	& --  & 1.16	& 1.11	& 1.08\\
 $\Delta E$/atom (eV)	& 9.56 &--	& 6.56	& 5.54 \\
	\hline
\end{tabular}
\caption{A comparison of graphene and \ssh versus graphone and {\gh} as reported 
in the literature.\cite{zhou_graphone,sofo:153401} $a$ is the lattice parameter, 
and $\Delta E$ is the binding energy.}
\label{tab:struct}
\end{table}

\begin{figure}
\begin{center}
\includegraphics[width=9cm]{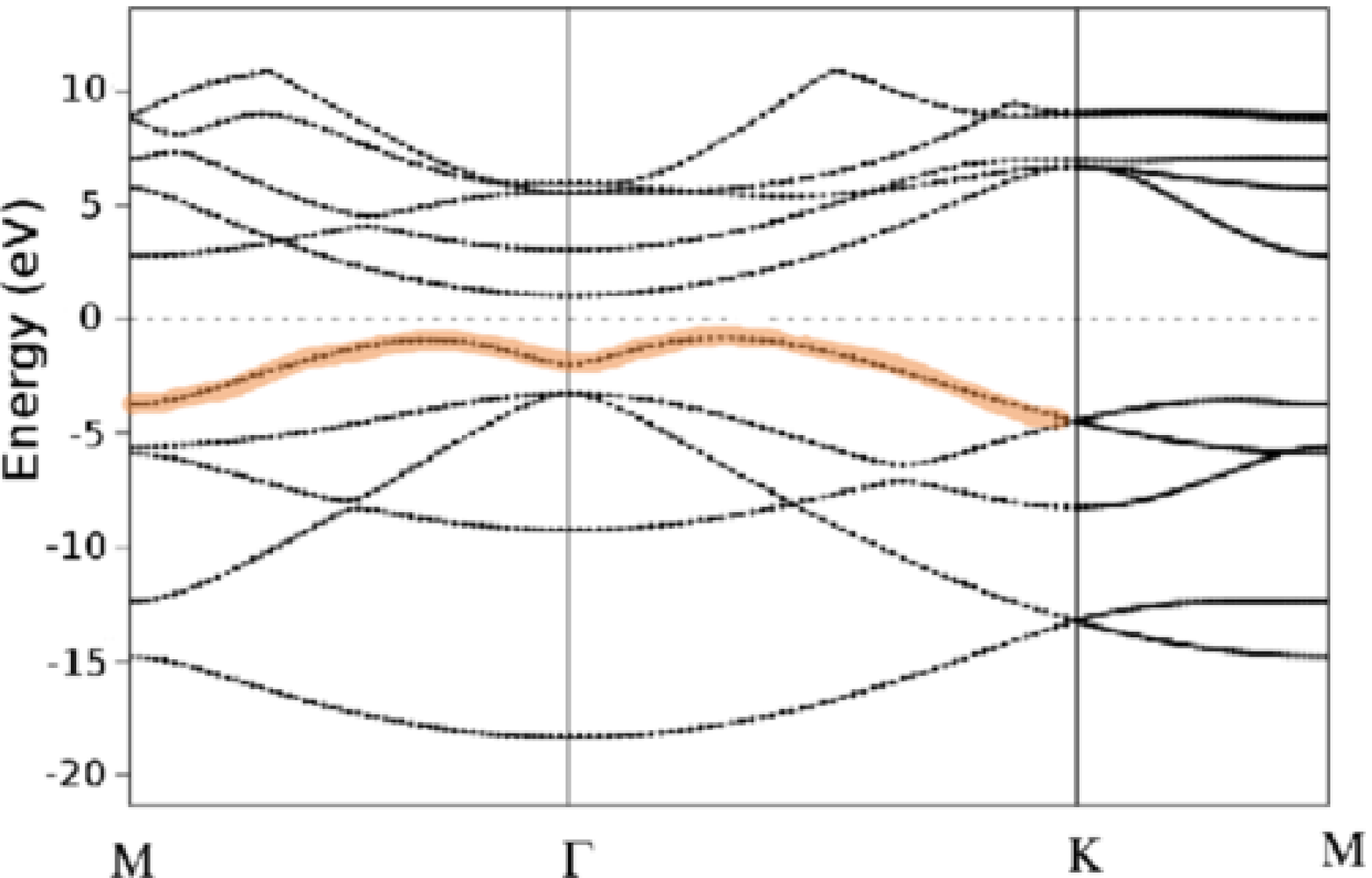}
\vskip 0.5cm
\includegraphics[width=9cm]{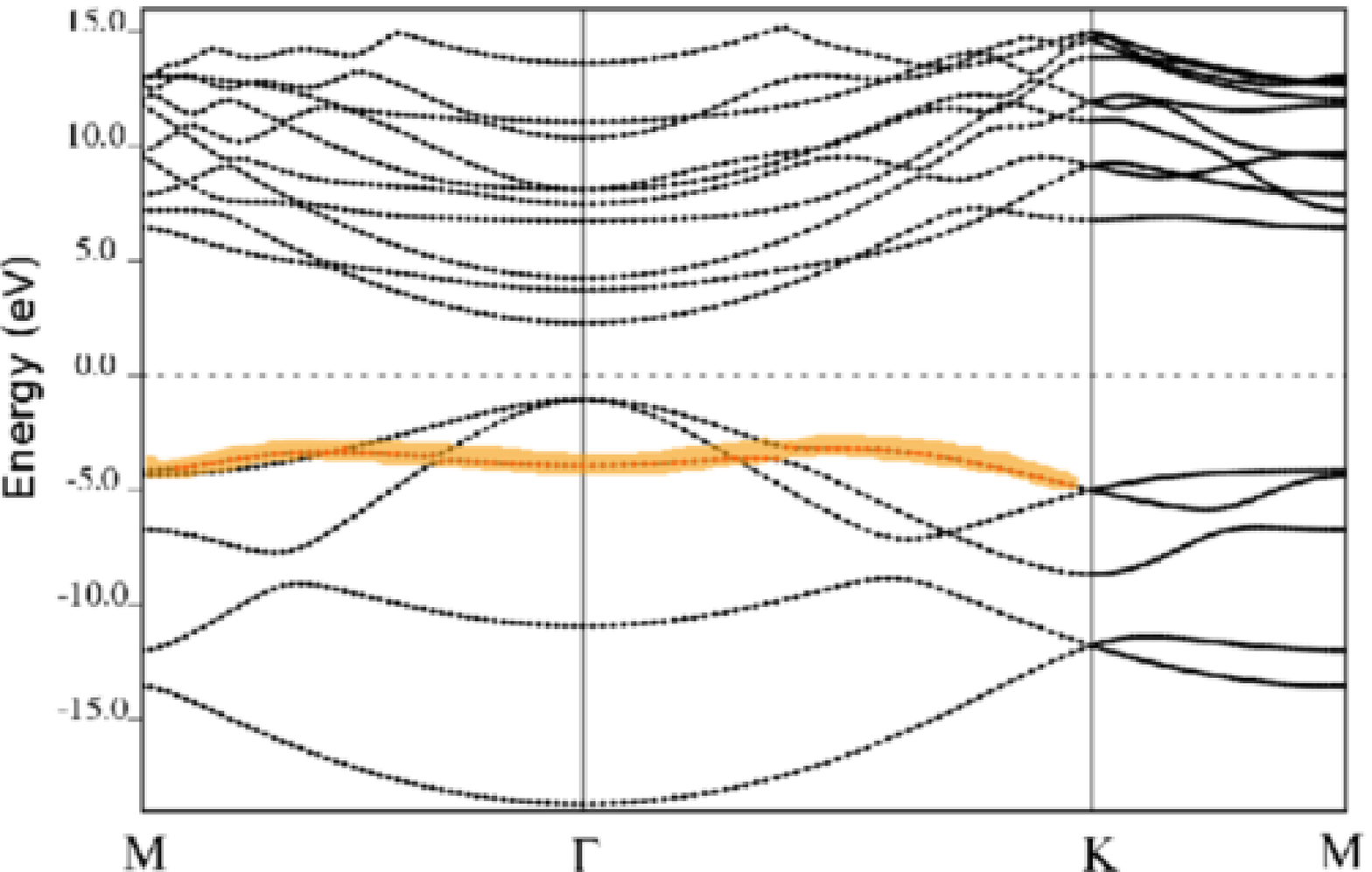}
\end{center}
\caption{Band structure of {\ssh} with the 1.89 eV indirect bandgap, versus {\gh}. 
Fermi level (dotted line) half way between the valance and conduction bands. 
The bold line marks the band which moved up in {\gh}, compared to {\ssh}.}
\label{fig:bs}
\end{figure}

\begin{figure}
\begin{center}
\includegraphics[width=8.5cm]{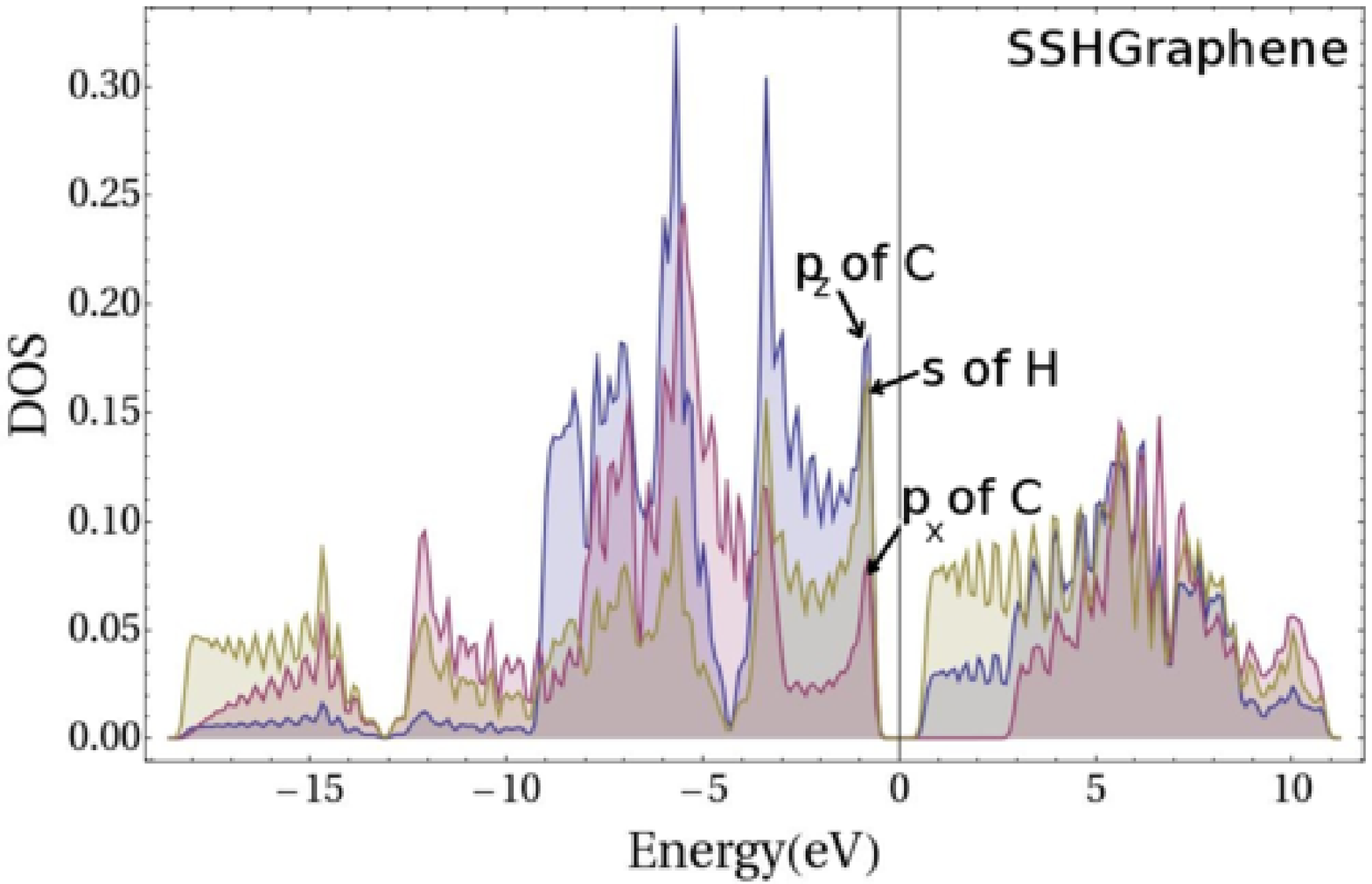}
\vskip 0.5cm
\includegraphics[width=8.5cm]{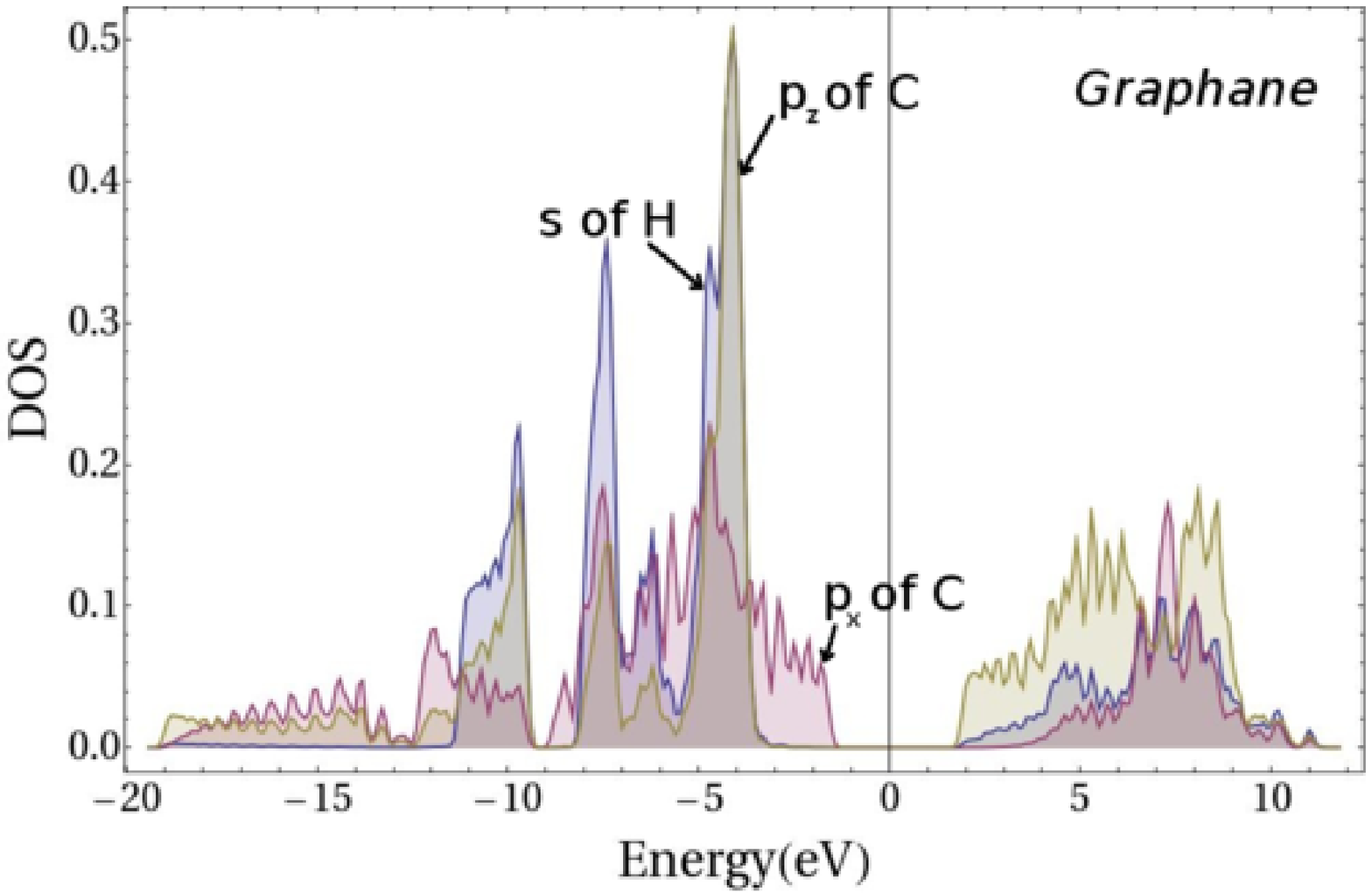}
\end{center}
\caption{Angular momentum decomposed DOS of {\ssh} and {\gh}.
$p_z$ from carbon and $s$ from hydrogen contribute substantially more than
$p_x$ and $p_y$ (identical to $p_x$, hence not shown). Fermi energy is taken
as zero. Gaussian broadening of 0.1 eV is applied for better visualization.}
\label{fig:pdos}
\end{figure}

We now approach the electronic structure of {\ssh}. It is well known that
the graphene band structure is very sensitive to deformations of any kind.
As noted before, there is a clear evidence that upon partial hydrogenation
the bandgap of graphene is opened. It is thus easy to conjecture that the
{\ssh} would be a semiconductor. However the most remarkable feature of
{\ssh} is that it is a semiconductor with an indirect bandgap. At this point
we recall that graphene is gapless, while {\gh} has a direct bandgap of 3.5 eV.

The band structure of {\ssh} shown in the upper part of Figure \ref{fig:bs} 
clearly exhibits an indirect bandgap of 1.89 eV. This is interesting, as this
value lies in between the gapless graphene and the rather wide bandgap {\gh}
(3.5 eV by DFT and 5.4 eV by GW method.\cite{gw}) Thus, {\ssh} becomes a preferred
organic candidate for semiconductor based devices. As mentioned earlier, a small
indirect bandgap was also seen by Zhou {\etal} \cite{zhou_graphone} in graphone.
However, we would like to point out that they did not report any optimization
of the unit cell, which may considerably affect the bandgap as well as lattice
parameters and bond lengths shown in Table \ref{tab:struct}.

The origin of the indirect bandgap can be inferred when we compare the band structures
of {\ssh} and {\gh} in Figure \ref{fig:bs}. A close examination reveals that the most
significant difference consists in the position of one band of {\gh} marked in orange.
In fact, this is the band that arises in graphene from the $p_z$ orbitals on carbon
atoms and is degenerate at the $K$ point. Upon hydrogenation this band lowers in energy
and the nature of hydrogenation determines the lowering.

In the case of {\gh} this band is significantly lower in energy as the hydrogenation 
is done in the best possible way, which is reflected in the stability seen in Table 
\ref{tab:struct}. On the other hand, in the single sided hydrogenation all the hydrogen
atoms are held at the same side, which naturally results in a repulsion among the hydrogen
atoms. This further causes the crystal to expand. As a result of enhanced repulsion among
the hydrogen atoms, the band marked in bold shifts upward in energy, which gives rise to
an indirect bandgap. In short, the peculiar band structure is due to the enhanced repulsion
among the hydrogen atoms which results in the shift of the most sensitive $p_z$ type band
of the system. This also suggests that we may expect the $p_z$ type character associated 
with this band in the case of {\ssh} as well. We verify the angular momentum character of 
the band by examining angular momentum decomposed density of states and by the partial 
charge density of the given band. Figure \ref{fig:pdos} presents the angular momentum 
decomposed DOS for {\ssh} and {\gh}, showing the $p_x$ (identical to $p_y$) and $p_z$ 
components of DOS on the carbon atom, along with the $s$ component on the hydrogen atom. 
It becomes evident from the comparison that in the case of {\ssh}, the $p_z$ of carbon 
atoms and $s$ of hydrogen atoms have shifted towards the Fermi level. In fact, the 
contribution from $p_z$ is almost twice that of $p_x$. This observation strengthens 
our earlier conjecture that the band moved upwards in energy.

\begin{figure}
\begin{center}
\includegraphics[width=8cm]{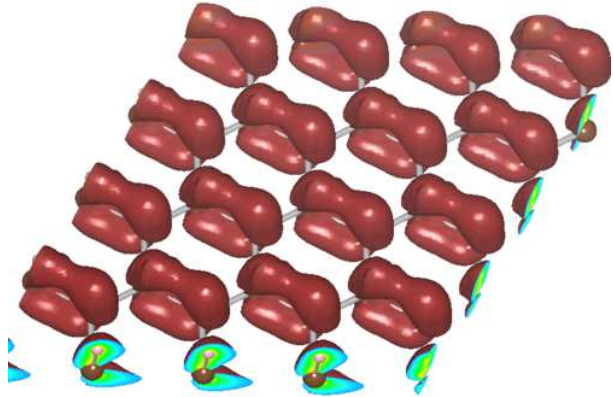}
\end{center}
\caption{Charge density isosurfaces of the band just below the Fermi level. 
The isosurface (at 0.1 of the maximum) covers the neighboring hydrogen atoms. 
Observe that the hydrogen atoms are connected along the arm-chair direction.}
\label{fig:iso} \end{figure}

The nature of DOS of {\ssh} also underlines a strong hybridization of $p_z$ of carbon 
atoms with $s$ of hydrogen atoms. This is the case, indeed, as seen from scrutinizing 
the individual bands. Figure \ref{fig:iso} shows the charge density isosurface of the 
top most occupied band. The density is plotted at the $\Gamma$ point. There are two 
clear observations to be drawn. First, there is a $\sigma$ bond established between 
the $p_z$ of carbon and $s$ of hydrogen. Second, there is an $s$-$s$ overlap between 
the two neighboring hydrogen atoms, established only along the arm-chair direction 
but not along the zig-zag.

At this point, it is worth noting that it may be inappropriate to call this covalently 
bonded system $sp^3$ hybridized, since the characteristic bond angle of 109.5$^\circ$ 
is not present anywhere. Generally in the case of a few hydrogen atoms interacting 
with graphene or even for {\gh}, the underlying carbon atoms are displaced from their 
locations, showing the signature of $sp^3$ bonded system. However, in {\ssh} all the 
carbon atoms remain in one plane, making it difficult to call it $sp^3$ hybridized.

In summary, we investigated the single sided hydrogenation of graphene using 
first-principle DFT, and demonstrated that upon complete relaxation of the unit 
cell and atomic degrees of freedom {\ssh} exhibits an indirect bandgap of 1.89 eV. 
This is appealing, as graphene is gapless while {\gh} is rather wide bandgap semiconductor, 
making {\ssh} a more desirable semiconductor. Our comparison of the band structures of 
{\ssh} and {\gh} shows that the $p_z$ band located in {\gh} significantly below the 
Fermi level moves up in {\ssh}, becoming the top most occupied band. This is attributed 
to an $s-s$ overlap between the hydrogen atoms. Our analysis of the binding energies 
clearly indicates stability of {\ssh}. Further accuracy check of the bandgap, bond 
lengths, and stability of {\ssh} we have predicted here can be addressed when 
experimental data will be available. \\

This work was supported by the Natural Sciences and Engineering Research Council (NSERC)
and National Research Council (NRC) of Canada. The research has been enabled by the use
of computing resources provided by WestGrid and Compute/Calcul Canada. B.S.P. also thanks
D. G. Kanhere, K. Joshi and D. Saraf for their valuable inputs. Some figures are generated
using VESTA.\cite{vesta}

%
%\bibliography{Pujari_etal_SSH_Graphene_Bibliography}
%

\end{document}

\title{{\Large Supplemental Material}\\ ~\\ 
Single-side-hydrogenated graphene: Density functional theory predictions}
\author{Bhalchandra S. Pujari}
\affiliation{National Institute for Nanotechnology, 11421 Saskatchewan Drive, Edmonton, AB, T6G 2M9, Canada}
\affiliation{Department of Mechanical Engineering, University of Alberta, Edmonton, AB, T6G 2G8, Canada}
\author{Sergey Gusarov}
\affiliation{National Institute for Nanotechnology, 11421 Saskatchewan Drive, Edmonton, AB, T6G 2M9, Canada}
\author{Michael Brett}
\affiliation{National Institute for Nanotechnology, 11421 Saskatchewan Drive, Edmonton, AB, T6G 2M9, Canada}
\affiliation{\hskip -1cm \centerline{Department of Electrical and Computer Engineering, University of Alberta, Edmonton, AB, T6G 2V4, Canada}}
\author{Andriy Kovalenko}
\altaffiliation{Corresponding Author: andriy.kovalenko@nrc-cnrc.gc.ca}
\affiliation{National Institute for Nanotechnology, 11421 Saskatchewan Drive, Edmonton, AB, T6G 2M9, Canada}
\affiliation{Department of Mechanical Engineering, University of Alberta, Edmonton, AB, T6G 2G8, Canada}

\maketitle

\subsection{Computational details}
\vskip -0.3cm

All the calculations are based on DFT using a plane wave basis as implemented in the
Quantum ESPRESSO package\cite{qe}. We use ultrasoft pseudopotentials \cite{Van} to
represent the interaction between ionic cores and valence electrons. The Generalized
Gradient Approximation (GGA) is used with the exchange-correlation functional of Perdew,
Burke and Ernzerhof (PBE).\cite{PBE} Kinetic energy cut-off on the plane wave basis for
wave functions is set to be 60 Ry ($\sim$ 816 eV) and for charge density is 240 Ry
($\sim$ 3265 eV) which are sufficiently large for graphene-like systems. A strict
convergence threshold for the self consistency is set at 10$^{-8}$ while that for the
forces is 0.005 eV/{\AA}. Further, to maintain the accuracy, integration over the
Brillouin zone is performed on regular 26 $\times$ 26 $\times$ 1 Monkhorst-Pack grids.
The band structure is plotted on the lines joining $M$, $\Gamma$, $K$ and $M$ points,
and the individual line segments are sampled using 50 grid points each. 
The higher level of precision is also maintained for the cell optimization which is
carried out using the BFGS quasi-Newton algorithm with Hellmann-Feynman forces and the
stress tensors. The convergence threshold on the pressure is kept at 0.1 kBar. The
computational unit cell consists of two carbon and two hydrogen atoms. A vacuum space
of 12 {\AA} is kept normal to the plane to avoid any interactions between the adjacent
sheets.

To verify that the features seen in the band structure are not artifacts of the
treatment on exchange-correlation functional, we have repeated the cell-optimization and
band structure calculations using a different treatment to exchange correlation
functionals including three different GGAs and one Local Density Approximation (LDA).
Details of the results are given in the following section. Here we stress that our
calculations confirmed that all are within reasonable agreement, except the bandgaps. This
is understandable as the bandgaps obtained using DFT are sensitive to the choice of the
functional. Since the qualitative picture remains unchanged, and since PBE is already
proven to be reliable with graphene, graphone and {\gh}, it is appropriate to report the
results obtained using GGA-PBE.

\vspace*{-0.5cm}
\subsection{Comparison of exchange-correlation functionals}
\vskip -0.3cm

Choice of exchange-correlation functional plays a crucial role in DFT calculations. To
test the accuracy of our result we now compare the results obtained using different
exchange correlation functionals. Except than the functional, other computational parameters
remain unchanged. We use three GGA based functionals, namely Perdew-Burke-Ernzerhof
\cite{PBE}, Heyd-Scuseria-Ernzerhof \cite{HSE,HSE2} and B3LYP \cite{b3lyp} as well as one
LDA based functional, Perdew-Zunger \cite{pz} for the comparison. The results obtained are
summarized in table \ref{tab:xc}.

Because within DFT formalism the electron density is physically meaningful and
the wave functions and energy bands are solutions of an eigenvalue problem for noninteracting electrons where their exchange and correlation interactions are
shifted into the functionals, the bandgap becomes sensitive to the choice of
functional. This is indeed seen in the Table (\ref{tab:xc}). Although the
bandgap values show considerable disagreement, other quantities remain in good
agreement with each other.

For a more conclusive value of the bandgap, treatment by a higher {\it ab initio} method is warranted.

\renewcommand{\baselinestretch}{1.3}
\begin{table}[!h]
	\centering
	\begin{tabular}{|c|c|c|c|c|}
		\hline
		~ 	& ~PBE\cite{PBE}~	& ~HSE\cite{HSE,HSE2}~ 	&
		~B3LYP\cite{b3lyp}~	& ~LDA\cite{pz}~ 	\\
		\hline
		Band Gap (eV)	& 1.89 	& 1.35	& 1.58	& 1.22 	\\
		Lattice Parameter (\AA)	& 2.84	& 2.82	& 2.87	& 2.80	\\
		C-C bond (\AA)		& 1.638	& 1.630	& 1.656	& 1.614 \\
		C-H bond (\AA)		& 1.076	& 1.085	& 1.070	& 1.075	\\
		\hline
	\end{tabular}
	\caption{Comparison of exchange-correlation	functionals.}
	\label{tab:xc}
\end{table}
\renewcommand{\baselinestretch}{1.0}

\bibliography{Pujari_etal_SSH_Graphene_Bibliography}